\documentclass[11pt,twoside]{article}

\usepackage{epsfig}
\usepackage{lscape}

\newcommand{\AAA}{\mbox{\boldmath $A$}}
\newcommand{\BB}{\mbox{\boldmath $B$}}
\newcommand{\JJ}{\mbox{\boldmath $J$}}
\newcommand{\bdot}{\mbox{\boldmath $\cdot$}}
\newcommand{\curl}{\mbox{\boldmath $\nabla \times$}}
\newcommand{\vort}{\mbox{\boldmath $\omega$}}
\newcommand{\vv}{\mbox{\boldmath $v$}}

\begin{document}
\title{Three-Dimensional Simulations of Solar and Stellar
Dynamos: The Influence of a Tachocline\footnote{To appear in Proc.\
GONG 2008/SOHO XXI Meeting on {\em Solar-Stellar Dynamos as 
Revealed by Helio and Asteroseismology}, held August 15-18, Boulder, CO,
Astronomical Society of the Pacific Conf.\ Ser., volume TBD.}}   
\author{Mark S. Miesch$^a$, Matthew K. Browning$^b$, A. Sacha Brun$^c$, \\
Juri Toomre$^d$, Benjamin P. Brown$^d$ \\
$^a$HAO/NCAR, Boulder, CO 80307-3000, USA \\
$^b$CITA, Univ.\ Toronto, 60 St.\ George St., Toronto, Ontario, M5S 3H8, Canada \\
$^c$DSM/DAPNIA/SAp, CEA Saclay, 91191 Gif-sur-Yvette, France \\
$^d$JILA, Dept.\ Astrophys.\ Planet.\ Sci., Univ.\ Colorado, Boulder, CO 80309-0440, USA}

\maketitle

\begin{abstract} 
 
We review recent advances in modeling global-scale convection and
dynamo processes with the Anelastic Spherical Harmonic (ASH) 
code.  In particular, we have recently achieved the first 
global-scale solar convection simulations that exhibit turbulent 
pumping of magnetic  flux into a simulated tachocline and the 
subsequent organization and amplification of toroidal field 
structures by rotational shear.   The presence of a tachocline
not only promotes the generation of mean toroidal flux, but
it also enhances and stabilizes the mean poloidal field throughout 
the convection zone, promoting dipolar structure with less frequent 
polarity reversals.  The magnetic field generated by a convective 
dynamo with a tachocline and overshoot region is also more helical
overall, with a sign reversal in the northern and southern hemispheres.  
Toroidal tachocline fields exhibit little indication of magnetic 
buoyancy instabilities but may be undergoing magneto-shear instabilities.
\end{abstract}



\section{Introduction}
There is little doubt that turbulent convection in the solar envelope must 
convert kinetic energy into magnetic energy by means of hydromagnetic dynamo 
action.  However, theoretical arguements and numerical simulations of 
turbulent dynamos suggest that much of this field must be in a disordered 
state, dominated by small-scale stochastic fluctuations.  The challenge of
global solar dynamo theory is to account for the ordered component of the
magnetic field, that which gives rise to sunspots, large-scale fields 
features such as dipolar and quadrupolar moments, and the 22-year 
activity cycle.

Advances in solar dynamo theory over the past several decades suggest that 
boundary layers may play a crucial role in generating this ordered magnetic
field component.  In particular, the solar tachocline located near the 
base of the convective envelope may be a breeding ground for the 
manufacture and accumulation of large-scale magnetic flux.  There 
are several reasons for this.  (I) Convective expulsion, turbulent
pumping, and global meridional circulations continually transport magnetic
field into the tachocline region where it can accumulate.  (II) Rotational
shear amplifies mean toroidal fields through the $\Omega$-effect and
suppresses smaller-scale fields by enhancing ohmic diffusion, thus
promoting the generation of strong, ordered toroidal flux.  (III) The 
subadiabatic stratification inhibits magnetic buoyancy instabilities, 
enabling long-term storage of magnetic flux.  (IV) The presence of strong,
stable large-scale toroidal flux and electrical current structures in 
the tachocline may provide a magnetic inertia, enhancing and stabilizing 
the poloidal field component as well.  Other processes may also contribute 
to the generation of ordered magnetic fields and cyclic activity, including 
the breakup of active regions in the upper boundary layer (the 
Babcock-Leighton mechanism) and advection of magnetic fields by global 
meridional circulations (e.g.\ Charbonneau 2005).

Here we review insight into the nature of global solar and stellar dynamo
processes gained from 3D MHD simulations of turbulent convection in 
rotating spherical shells.  We focus on two representative dynamo simulations
in a solar context, one in which the computational domain is limited to the 
convective envelope and one that incorporates convective overshoot into an 
underlying radiative zone where uniform rotation is imposed, thus creating 
a mock tachocline.  In \S2-3 we compare the two simulations and assess how 
the presence of this tachocline and overshoot region alters the global dynamo.  
The most apparent difference is the presence of strong, organized toroidal 
fields in the tachocline region of the latter simulation.  In \S4 we discuss 
how such fields may be maintained and address their stability and 
evolution.  In \S5 we briefly summarize the implications of these and
other stellar dynamo simulations.

\section{Simulations and Flow Features}
In order to assess the role of a tachocline in global dynamo action
we consider two 3D MHD simulations of global solar convection.  Both
are sustained dynamos that extend over multiple ohmic diffusion time
scales.  The first is described in detail by Brun, Miesch \& Toomre 
(2004; hereafter BMT04), and is referred to there as Case M3.  Here we refer to 
it as Case A.  The second, which we refer to here as Case B, is described 
further by Browning et al.\ (2006; hereafter BMBT06; see 
also Browning et al.\ 2007).  Both simulations are based on the
Anelastic Spherical Harmonic (ASH) code with solar values used for
the luminosity, rotation rate, and background stratification.

\begin{figure}[t!]
\epsfig{file=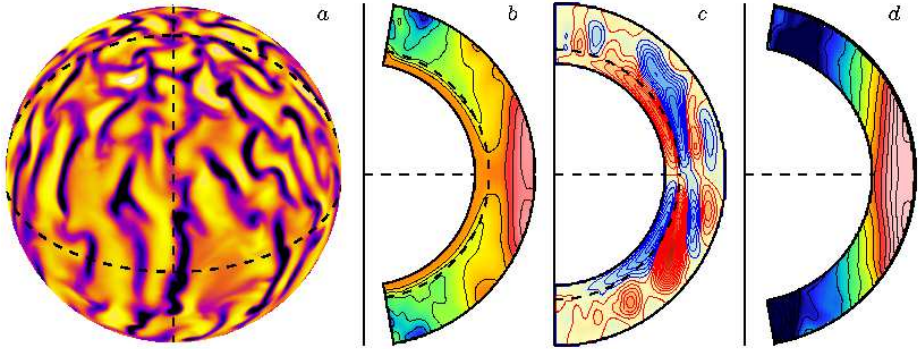,width=\linewidth}
\caption{($a$-$c$) Illustration of convective patterns and mean flows in Case B.
($a$) Orthographic projection of the radial velocity, $v_r$ near the
middle of the convection zone ($r = 0.84 R$), with the north pole titled
35$^\circ$ toward the line of sight.  Dashed lines indicate the equator and
several meridians.  Bright tones (yellow/orange in online version) denote 
upflow and dark tones (blue/black) denote downflow.  ($b$) Angular velocity 
and ($c$) meridional circulation averaged over longitude and time (90 days), 
the latter represented as streamlines of the mass flux.  The dashed line 
indicates the base of the convection zone.  ($d$) The angular velocity in Case A
is also shown for comparison.  Bright (pink/red) and dark (blue/black) tones
denote faster and slower rotation in ($b$) and ($d$), ranging from 
350 to 450 nHz.  Bright (red) and dark (blue) tones in ($c$) denote
clockwise and counter-clockwise circulations respectively, with the
flow speed ranging from 2-20 m s$^{-1}$.\label{drmc}}
\end{figure}

The principle difference between Cases A and B is the presence of a
tachocline in the latter.  Whereas the computational domain in
Case A stops at the base of the convection zone, $r = 0.72 R$ where $R$
is the solar radius, Case B incorporates a portion of the subadiabatic 
radiative interior, extending downward to $r = 0.62 R$.  Both simulations 
stop short of the solar photosphere, with upper boundaries at (A) $0.97 R$ 
and (B) $0.96 R$, in order to avoid complications in the photospheric 
boundary layer such as granulation, ionization, and the transition to 
radiative energy transfer which cannot be easily accounted for in our 
global anelastic modeling framework.  The kinematic viscosity, $\nu$, 
is somewhat higher in Case B, $2\times10^{12}$ cm$^2$ s$^{-1}$ in the
mid convection zone versus $1.4\times10^{12}$ cm$^2$ s$^{-1}$ in Case A,
but the magnetic diffusivity $\eta$ is somewhat lower 
($P_m = 4$ in Case A and $P_m = 8$ in Case B where $P_m = \nu / \eta$).
Both simulations have comparable spatial resolution.  For further
details see BMT04 and BMBT06.

A tachocline is imposed in Case B by means of a drag force in the radiative
interior that imposes uniform rotation.  This is intended to take into
account tachocline confinement processes that are not captured by
the simulation due to limited spatial resolution and temporal coverage.
A thermal forcing is also applied near the base of the convection zone,
warming the poles relative to the equator.  This is intended to help
sustain a strong rotational shear across the tachocline by means of
baroclinic torques that partially offset the sub-grid-scale diffusion
and that compensate for the extent of the overshoot region which is wider
than in the Sun due to computational limitations.  We emphasize 
that the rotational shear in the convective envelope is still maintained 
by the convective Reynolds stress; the mechanical and thermal
forcing merely promotes a sharp tachocline.  Such an approach cannot
provide much insight into tachocline confinement processes but it can
be used to investigate the influence of a tachocline on global
dynamo action, which is our motivation.  For further discussion 
see BMBT06.

The convective structure in Cases A and B is similar, dominated at low
latitudes in the mid convection zone by columnar convection cells
aligned with the rotation axis (Fig.\ \ref{drmc}$a$), as in comparable
non-magnetic simulations (reviewed by Miesch \& Toomre 2009).  Near the 
surface both exhibit an interconnected network of downflow lanes laced by 
vertical magnetic flux (BMT04). 

The differential rotation profile in 
both cases is roughly solar-like (Fig.\ \ref{drmc}$b$, $d$), with a 
monotonic decrease in angular velocity $\Omega$ with latitude.  However,
both exhibit more cylindrical alignment than the solar rotation
profile as revealed by helioseismology.  Furthermore, the angular
velocity contrast $\Delta \Omega$ between equator and pole is much
weaker in Case B (about 13\% compared to 36\% in Case A).  This is typical 
of penetrative convection simulations and may arise in part from our 
artificially wide overshoot region (Miesch 2007a) and in part from Lorentz 
forces associated with stronger mean fields (\S3).  Despite the 
weaker $\Delta \Omega$, Case B possesses a tachocline; substantial radial 
shear near the base of the convection zone maintained by our mechanical
and thermal forcing.

The most prominent feature of the mean meridional circulation in Case B is
an equatorward flow at the base of the convection zone with an amplitude 
of about 8 m s$^{-1}$ (Fig.\ \ref{drmc}$c$).  This may in part be attributed 
to the thermal forcing but other simulations without such forcing also exhibit 
equatorward circulation near the base of the convection zone, albeit somewhat 
weaker (several m s$^{-1}$; see Miesch et al.\ 2000).  This has important implications
for flux-transport dynamo models as we address in \S4.  The circulation profile 
in the bulk of the convection zone is multi-celled, as in Case A (BMT04).

\begin{figure}[t!]
\epsfig{file=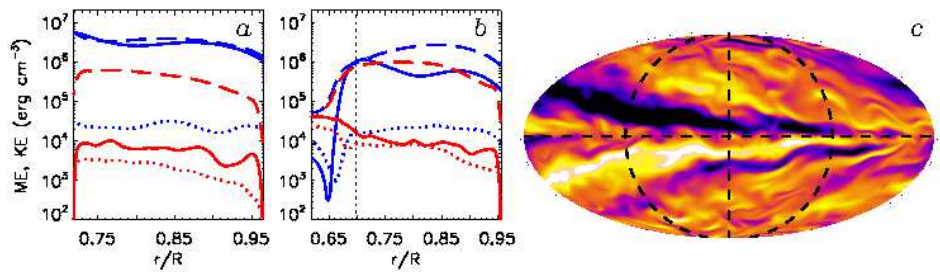,width=\linewidth}
\caption{Magentic (black; red in online version) and kinetic (grey; blue) energy density in ($a$) 
Case A and ($b$) Case B, averaged over horizontal surfaces and over time (4 days).  Solid lines denote 
the energy in the mean
toroidal field and the differential rotation, dotted lines that in the mean poloidal field and the meridional
circulation, and dashed lines that in the non-axisymmetric field and flow components.  The vertical dotted
line in ($b$) denotes the base of the convection zone.  ($c$) Molleweide projection of the longitudinal
field component $B_\phi$ in the tachocline of Case B ($r=0.67R$) at a time when $m=1$ structure is 
particularly evident.  Bright and dark tones denote eastward and westward fields respectively.\label{energetics}}
\end{figure}

\section{Mean and Fluctuating Fields}
As discussed in \S1, turbulent flows beget turbulent fields.  In both Cases A and B, dynamo 
action produces strong fluctuating (non-axisymmetric) fields that account for 95-98\% of 
the total magnetic energy in the convection zone (Fig.\ \ref{energetics}$a$, $b$).  Both
mean and fluctuating field components are somewhat stronger in Case B, due in part to
the lower magnetic diffusivity (\S2).   More remarkably, the magnetic energy in the
mean toroidal field of Case B rises steadily toward the base of the convection zone, 
peaking in the tachocline where it reaches equipartition with the fluctuating field
components and also with the fluctuating kinetic energy.  Note that the drop in the
differential rotation kinetic energy below $r = 0.69R$ is due primarily to the 
mechanical forcing (\S2).

\begin{figure}[t!]
\epsfig{file=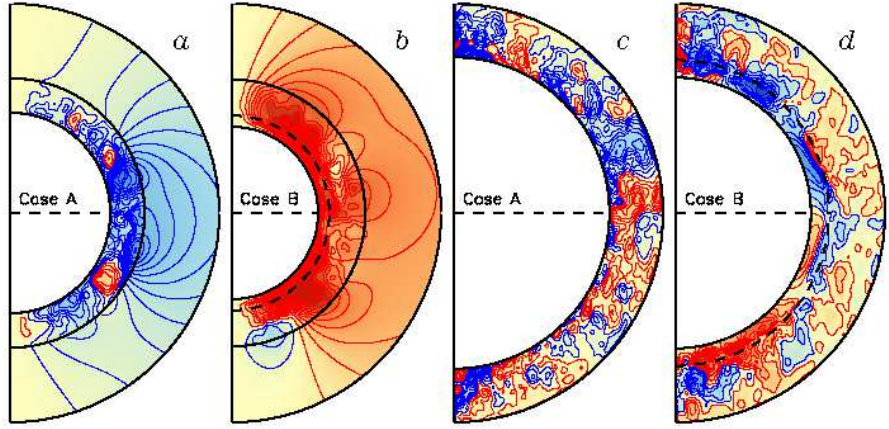,width=\linewidth}
\caption{Mean ($a$, $b$) poloidal magnetic fields and ($c$, $d$) toroidal magnetic fields in
($a$, $c$) Case A and ($b$, $d$) Case B, averaged over longitude and time (90 days).  
White (red in online version) tones denote a clockwise poloidal field orientation and
eastward toroidal field whereas black (blue) tones denote the opposite.  Saturation levels
for the poloidal and toroidal color tables are $\pm$20G and $\pm$2000G respectively.  
Poloidal fields in ($a$, $b$) include a potential-field extrapolation out to $1.5R$ 
and the base of the convection zone is indicated by a dashed line 
in ($b$, $d$).\label{Bmean}}
\end{figure}

The structure of the tachocline field in Case B is dominated by strong
(2-5 kG) toroidal bands or extended magnetic layers that are roughly
antisymmetric about the equator.  These toroidal bands are generally 
axisymmetric as illustrated in Figure \ref{Bmean}$d$ but they occasionally 
exhibit a pronounced $m=1$ component as demonstrated in Figure 
\ref{energetics}$c$,  where $m$ is the longitudinal wavenumber.  This 
$m=1$ behavior may be indicative of magneto-shear instabilities as we 
discuss in \S4.  The toroidal bands are remarkably persistent, remaining 
intact for thousands of days, punctuated sporadically by brief 
intervals ($<$ 200 days) of roughly symmetric parity about the equator 
after which the antisymmetric parity is re-established with the same
polarity as before (BMBT06, Browning et al.\ 2007).  The simulation
has spanned over two decades and has not yet undergone a sustained
polarity reversal of the mean toroidal field in the tachocline.

\begin{figure}[t!]
\epsfig{file=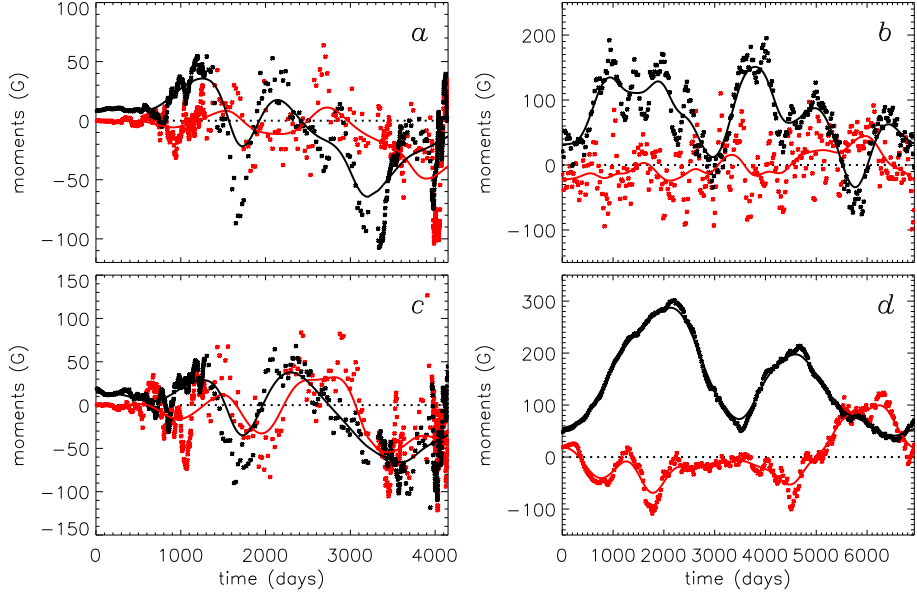,width=\linewidth}
\caption{Time evolution of the dipole (black) and quadrupole (grey; red in online version) 
moments in ($a$, $c$) Case A and ($b$, $d$) Case B.  The top row corresponds to the
outer boundary in each case ($a$) $0.97R$ and ($b$) $0.96R$ whereas the bottom
row corresponds to the base of the convection zone in Case A ($c$) $0.72R$ and to the 
underlying tachocline in Case B ($d$) $0.67R$.  Points denote instantaneous snapshots
and lines denote running six-month averages.\label{dmom}}
\end{figure}

In contrast to the strong, persistent bands in the tachocline, the mean 
toroidal field in the convection zone of both cases A and B is relatively 
weak ($\sim$ 1 kG), disordered, and transient (Fig.\ \ref{Bmean}$c$, $d$).  
This lends support to the arguments put forth in \S1 on why the tachocline 
might be a prime place to generate large-scale magnetic flux.  The mean 
poloidal field in Case B (Fig.\ \ref{Bmean}$b$) also appears stronger and
more ordered than in Case A (Fig.\ \ref{Bmean}$a$), with a more 
prominent dipole moment.  This is demonstrated further in 
Figure \ref{dmom} which shows the dipole and quadropole moment
in each case over thousands of days of evolution.

In Case A, the dipole and quadrupole moments are comparable in amplitude
and erratically reverse sign on a time scale of roughly 500 days.  By
contrast, the dipole moment in Case B is stronger both in absolute 
magnitude and relative to the quadrupole component, although this
is less so toward the end of the interval shown in Figure \ref{dmom}.
The dipole moment is also more stable in Case B, maintaining the 
same sign through most of the simulation apart from a brief reversal
near the top of the shell at $t \sim $ 6000 days.  As with any
numerical simulation, however, these conclusions are based on
a limited time interval.  Figure \ref{dmom}$d$ in particular
suggests that the dipole moment may be waning over the course
of several decades.  Further time evolution is necessary to clarify
any potential long-term trends, including the possiblity 
of systematic polarity reversals.

\begin{figure}[t!]
\epsfig{file=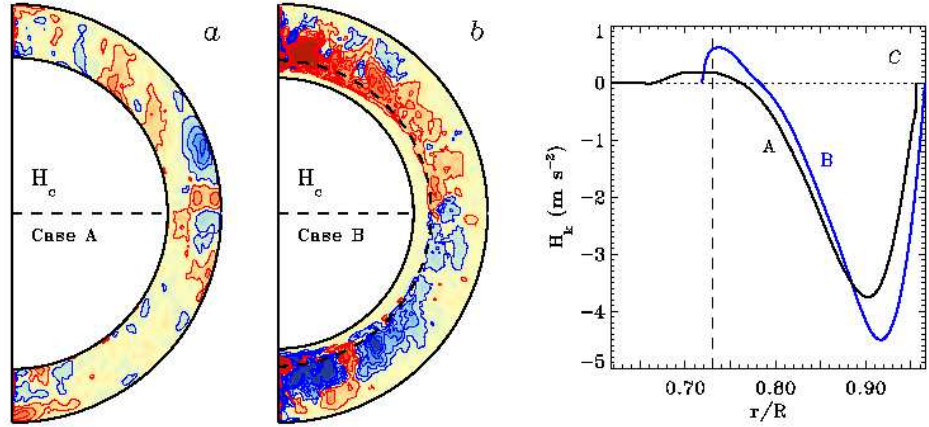,width=\linewidth}
\caption{The fluctuating current helicity $\JJ^\prime \bdot \BB^\prime$ is shown for
($a$) Case A and ($b$) Case B averaged over longitude and over the same 90-day time
interval as in Figures \ref{drmc} and \ref{Bmean}. White and black (red and blue in 
online version) denote positive and negative values and the color table saturates
at $\pm 2\times 10^7$ statamp cm$^{-2}$ G. ($c$) Kinetic helicity in Case A (black
line) and Case B (grey; blue), averaged over the northern hemisphere and time
(4 days).  Dashed lines in ($b$) and ($c$) denote the base of the convection zone
in Case B.\label{helicity}}
\end{figure}

Another remarkable feature of the magnetic field in Case B is illustrated
in Figure \ref{helicity}$b$; it is helical.  Shown is the current helicity
in the fluctuating field component, $H_c^\prime = \JJ^\prime \bdot \BB^\prime$,
where $\JJ  =(c/4\pi) \curl \BB$ is the current density and primes indicate
that the longitudinal mean has been subtracted out.  Although the
magnetic helicity $H_m = \AAA \bdot \BB$ is of greater theoretical interest
(where $\AAA$ is the vector potential, defined such that $\BB = \curl \AAA$),
$H_c$ has the advantage that it is guauge invariant and is more readily
observable.  Both $H_c$ and $H_m$ quantify the degree to which the magnetic 
field topology is twisted and is not reflectionally symmetric.

Magnetic helicity has profound implications for large-scale dynamo action.
The turbulent $\alpha$-effect, whereby large-scale fields are generated
by small-scale velocity fields, is intimately tied to the upscale 
spectral transfer of magnetic helicity.  If the transfer is local, 
this is referred to as an inverse cascade.  However, the ensuing
buildup of small-scale helicity can suppress the inverse cascade
if it is not dissipated or removed through the boundaries.
This suppression can be so dramatic that it has been referred
to as catastrophic $\alpha$-quenching and if it occurs, it 
would imply that the turbulent $\alpha$-effect cannot account
for the mean fields observed in the Sun and other stars.
For further discussion see Brandenburg \& Subramanian (2005) 
and Miesch \& Toomre (2009). 

The helical nature of the fluctuating magnetic field in 
Case B (Fig.\ \ref{helicity}$b$) suggests that the presence 
of a tachocline and overshoot region may mitigate 
catastrophic $\alpha$ quenching.  The convection may be 
shunting small-scale helicity into the stable zone
where it is stored or dissipated, thus promoting 
helicity generation, and possibly self-organization processes
associated with upscale helicity transfer, in the convection 
zone.  By contrast, the fluctuating field in 
Case A (Fig.\ \ref{helicity}$a$) is not particularly helical 
and does not exhibit the asymmetry about the equatorial 
plane evident in Case B.

The sign of the current helicity in Case B is positive in the
north and negative in the south, which is opposite to that
inferred from photospheric and coronal measurements (e.g.\
Pevtsov \& Balasubramaniam 2003).  However, such measurements
are concerned with field structures such as active regions
that are associated with the large-scale field component.
If such fields are generated through an upscale transfer of 
magnetic helicity, one would expect that the sense of the 
helicity should be opposite to that in the fluctuating
component.  Indeed, the negative $B_\phi$ apparent in the 
northern hemisphere in Fig.\ \ref{Bmean}$d$ coupled with
the positive dipole moment apparent in Fig.\ \ref{Bmean}$d$
implies a negative current helicity for the mean field
component.  Thus, the sign of the current helicity in 
Case B may be consistent with solar observations.

There are many caveats to this simplified picture, including
the poorly understood details of how helicial flux tubes
form in the tachocline and emerge.   Nevertheless, it is 
clear that the dynamical aspects of magnetic helicity 
offer a fertile ground for further insight into the 
subtleties of global solar and stellar dynamos, 
particularly with regard to influence of a tachocline.

In contrast to $H_c^\prime$, the kinetic helicity, 
$H_k = \vort \bdot \vv$ where $\vv$ and $\vort$ are the fluid
velocity and vorticity, is similar in Cases A and B (Fig.\ \ref{helicity}$c$). 
The $H_k$ profile in both cases is antisymmetric about the equator, with 
a negative sign in the northern hemisphere through most of the convection 
zone.  Near the base of the convection zone the sign of $H_k$
reverses due largely to the influence of the Coriolis force on 
diverging downflows.  Kinetic helicity profiles such as this
are a well known feature of rotating compressible convection
(e.g.\ Gilman 1983; Miesch et al.\ 2000).  

\section{Maintenance and Evolution of Tachocline Fields}
\label{tachocline}

According to our current understanding of the solar dynamo, we may
expect that toroidal fields generated in the tachocline by 
the $\Omega$-effect should migrate equatorward, intermittently 
spawning flux tubes as they go which then emerge through the 
photosphere as active regions.  In this section we briefly address 
how our simulations relate to this prevailing dynamo paradigm.

To this end, we must first understand how the toroidal fields
in the tachocline of Case B are established and maintained.
A detailed analysis will await a future paper but here we can 
confirm that the principle source term in generating mean 
toroidal flux in the tachocline is indeed the $\Omega$-effect 
arising from radial angular velocity shear $\partial \Omega/\partial r$, 
as might be expected based on the prevailing paradigm.  The latitudinal 
shear $\partial \Omega/\partial\theta$ also contributes,
as does the radial component of the turbulent emf:
$E_r^\prime = v_\theta^\prime B_\phi^\prime - v_\phi^\prime B_\theta^\prime$.

These generation terms work to offset radial diffusion and the latitudinal 
advection of toroidal flux by the meridional circulation which both act
to erode the tachocline field (latitudinal diffusion is negligible).  
The latitudinal component of the turbulent emf,
$E_\theta^\prime = v_\phi^\prime B_r^\prime - v_r^\prime B_\phi^\prime$.
also acts as a sink term on average, at least over the 6930-day time 
interval shown in Fig.\ \ref{dmom}$b$, $d$.  This is somewhat 
surprising since turbulent pumping of toroidal flux, if it occurs,
would be reflected in the $E_\theta^\prime$ term.  Thus, turbulent
pumping acts to erode the tachocline field in Case B rather than
maintain it.  Although this may be contrary to theoretical expectations,
it is still consistent with the concept of turbulent pumping since the
mean toroidal flux in the convection zone of Case B is opposite in
sign (albeit weaker in magnitude) relative to the tachocline field.

Conventional wisdom suggests that as toroidal flux is amplified by
rotational shear in the tachocline it will eventually become 
susceptible to magnetic buoyancy instabilies, forming isolated
flux tubes which then rise toward the solar surface.  Although
some strong, transient toroidal ribbons in Case B do exhibit
a slight density deficit which may induce them to rise, the
persisent magnetic layer evident in Figure \ref{Bmean}$d$
appears to be largely stable.  The proximate reason for this
is because the layer does not exhibit any significant density 
evacuation relative to its surroundings but the ultimate 
explanation is still under investigation.  It may
be that our subgrid-scale diffusion and our artificially 
wide overshoot region limits the radial magnetic field gradients
that can be achieved.  In any case, it is clear that turbulent 
flows and fields in Case B contribute to the mechanical equilibrium
of the magnetic layer and furthermore, the layer is not an 
isolated magnetic surface; rather, it has field line connectivity
to the entire convection zone.  Thus, conditions here are 
far from the idealized equilibrium states typically considered
in investigations of magnetic buoyancy instabilities.

Toroidal magnetic fields in the tachocline are also susceptible
to another type of instability fed by latitudinal shear, as 
described by Gilman \& Fox (1997).  Although higher wavenumber
modes may also occur, the most vigorous and robust mode for
broad toroidal field profiles (antisymmetric about the 
equatorial plane) is the $m=1$ clam-shell instability whereby 
toroidal loops tip out of phase, reconnecting across the 
equator at one meridian and spreading poleward at the antipode.
If the rotational shear is maintained and if the poloidal 
field is continually replenished, such clam-shell instabilities
can occur indefinitely (Miesch 2007b).

The presence of sustained clam-shell instabilities in Case B is 
suggested by the prominent $m=1$ structure in snapshots such
as that shown in Fig.\ \ref{energetics}$c$.  If such instabilites 
are indeed occurring, one would expect to see a quasi-periodic
energy exchange between the $m=0$ and the $m=1$ components of
the toroidal field (Miesch 2007b).  We will address this 
in a future paper.

In mean-field dynamo models, the equatorward propagation of
toroidal flux implied by the solar butterfly diagram is
usually attributed to one of two mechanisms (that are not
mutually exclusive).  The first is advection by an equatorward
meridional circulation near the base of the convection zone
and the second is the propagation of a dynamo wave induced
by rotational shear and turbulent field generation
(the $\alpha$-effect).  In the latter case, 
equatorward propagation is achieved if the product
$H_k \partial \Omega / \partial r > 0$ in the northern
hemisphere (e.g.\ Charbonneau 2005).

Both of these propagation mechanisms could in principle be
occurring in Case B.  At mid and low latitudes near the base of the 
convection zone there is a strong equatorward meridional circulation 
(Fig.\ \ref{drmc}$c$) and $H_k$ and $\partial \Omega / \partial r$ are 
both positive in the northern hemisphere (Figs.\ \ref{drmc}$b$, 
\ref{helicity}$c$).  However, the amplitude of the circulation
and the kinetic helicity peak above the principle toroidal flux
concentration.  Perhaps not surprisingly, then, there is little
indication for any latitudinal propagation of the tacholine
field in Case B.

\section{Conclusion}
\label{conclusion}

Although 3D MHD dynamo simulations cannot capture all processes of
relevance to the global solar dynamo, they can provide crucial
insight into key ingredients.  In particular, understanding the 
subtle interaction between turbulent field generation in the 
convection zone and the generation of toroidal flux by rotational 
shear in the tachocline stands to benefit greatly from high-resolution
MHD simulations of penetrative convection.  By providing a reservoir 
for expelling, amplifying and storing magnetic flux, the tachocline 
influences the global behavior of the dynamo, promoting strong, 
stable mean fields and helical magnetic topologies throughout 
the convection zone.

Throughout this brief paper our emphasis has been on the Sun but 
stellar dynamo simulations have also produced substaintial insights
in recent years.  Highlights include the generation of strong toroidal 
flux structures in rapidly-rotating solar-like stars (Brown et al.\ 2007), 
enhancement of core dynamo action in A stars by fossil envelope fields 
(Featherstone et al.\ 2007), quenching of $\Delta \Omega$ by the 
Lorentz force in the deep convective shells of A and M stars
(Brun, Browning \& Toomre 2005; Browning 2008), and the saturation
of $\Delta \Omega$ and mean field strengths at high $\Omega$
(Christensen \& Aubert 2006).  The latter was done in planetary
context but may also have important implications for stars (see
Miesch \& Toomre 2009).

\vspace{.1in}
We thank Nicholas Featherstone, Kyle Augustson and Nicholas Nelson for 
numerous discussions on all aspects of MHD dynamo simulations and Keith 
MacGregor for helpful comments on the manuscript.  The work 
presented here was supported by NASA through the Heliophysics Theory 
Program grant NNG05G124G. The simulations were carried out with NSF 
PACI support of PSC, SDSC, NCSA, NASA support of Project Columbia, and
through the CEA resource of CCRT and CNRS-IDRIS in France.

\end{document}